\documentclass[10pt]{article}

\usepackage{amsmath}
\usepackage{amssymb}
\usepackage{amsthm}
\usepackage{amsfonts}

\usepackage{graphicx, subfig}

\usepackage{cite}

\usepackage{hyperref}

\usepackage{lineno}

\usepackage{microtype}
\DisableLigatures[f]{encoding = *, family = * }

\usepackage{color} 
\usepackage{epstopdf}

\usepackage{url}

\usepackage[labelfont=bf,labelsep=period,justification=raggedright]{caption}

\makeatletter
\renewcommand{\@biblabel}[1]{\quad#1.}
\makeatother

\date{}

\usepackage{float}
\usepackage{natbib}
\usepackage{xspace}
\usepackage{algorithm}
\usepackage{algorithmic}
\usepackage{tikz}
\usepackage{booktabs,ctable}
\newenvironment{methods}{}

\begin{document}
\newtheorem{property}{Property}
\newtheorem{definition}{Definition}
\newcommand{\avatar}{{\sc Avatar}\xspace}
\newcommand{\boolnet}{{\sc BoolNet}\xspace}
\newcommand{\firefront}{{\sc Firefront}\xspace}
\setlength{\heavyrulewidth}{0.2em}
\newcommand{\otoprule}{\midrule[\heavyrulewidth]}
\newcommand{\textcolon}{:}

\newcommand{\indic}{\ensuremath{1\!\!1}}
\newcommand{\nat}{\ensuremath{\mathbb{N}}}

\renewcommand{\AA}{\ensuremath{\mathcal{A}}}
\newcommand{\CC}{\ensuremath{\mathcal{C}}}
\newcommand{\DD}{\ensuremath{\mathcal{D}}}
\newcommand{\II}{\ensuremath{\mathcal{I}}}
\newcommand{\KK}{\ensuremath{\mathcal{K}}}
\newcommand{\TT}{\ensuremath{\mathcal{T}}}
\newcommand{\EE}{\ensuremath{\mathcal{E}}}
\newcommand{\FF}{\ensuremath{\mathcal{F}}}
\newcommand{\GG}{\ensuremath{\mathcal{G}}}
\newcommand{\MM}{\ensuremath{\mathcal{M}}}
\newcommand{\RR}{\ensuremath{\mathcal{R}}}
\newcommand{\cS}{\ensuremath{\mathcal{S}}}
\newcommand{\cV}{\ensuremath{\mathcal{V}}}
\renewcommand{\KK}{\ensuremath{\mathcal{K}}}

\newcommand{\Cr}{\ensuremath{C''}}
\newcommand{\Succ}[1]{\ensuremath{\mbox{\rm Succ}(#1)}\xspace}

\renewcommand{\P}{\mathbb{P}}

\newcommand{\transition}[3]{(#1, #2)}
\newcommand{\Max}{\mbox{\rm Max}}
\newcommand{\versor}[1]{\mathbf{e}^{#1}}
\newcommand{\reg}[2]{\mbox{\rm Reg}_{#2}(#1)}
\newcommand{\map}[2]{\mbox{\rm Map}_{#1}^{(#2)}}

\begin{flushleft}
{\Large
\textbf{Quantification of reachable attractors in asynchronous discrete dynamics}
}
\\
Nuno D. Mendes\,$^{1,3,4,\dagger}$,
Pedro T. Monteiro\,$^{1,3,\dagger}$,
Jorge Carneiro\,$^{1}$,
Elisabeth Remy\,$^{2}$,
Claudine Chaouiya\,$^{1,\ast}$
\\
\bf{1} Instituto Gulbenkian de Ci\^encia, Oeiras, Portugal
\\
\bf{2} Institut de Math\'ematiques de Luminy, Marseille, France
\\
\bf{3} Instituto de Engenharia de Sistemas e Computadores - Investiga\c{c}\~ao e Desenvolvimento (INESC-ID), Lisbon, Portugal
\\
\bf{4} IBET, Oeiras, Portugal
\\
{\footnotesize $\ast$ E-mail: chaouiya@igc.gulbenkian.pt}\\
{\footnotesize $\dagger$ these authors contributed equally to this work.}
\end{flushleft}

\section*{Abstract}
{\bf Motivation:}
Models of discrete concurrent systems often lead to huge and complex state transition graphs that represent their dynamics. This makes difficult to analyse dynamical properties. In particular, for logical models of biological regulatory networks, it is of real interest to study attractors and their reachability from specific initial conditions, i.e. to assess the potential asymptotical behaviours of the system. Beyond the identification of the reachable attractors, we propose to quantify this reachability.

\noindent
{\bf Results:}
Relying on the structure of the state transition graph, we estimate the probability of each attractor reachable from a given initial condition or from a portion of the state space. First, we present a quasi-exact solution with an original algorithm called \firefront, based on the exhaustive exploration of the reachable state space. Then, we introduce an adapted version of Monte Carlo simulation algorithm, termed \avatar, better suited to larger models. \firefront and \avatar methods are validated and compared to other related approaches, using as test cases logical models of synthetic and biological networks.

\noindent
{\bf Availability:}
Both algorithms  are implemented as Perl scripts that can be freely downloaded from \url{http://compbio.igc.gulbenkian.pt/nmd/node/59} along with Supplementary Material.

\section{Introduction}

The logical modelling framework has been widely used to study gene regulatory and signalling networks~(see e.g. \citep{saadatpour2012,glass2010}). Briefly, in these models, discretised levels of the components evolve depending on the levels of their regulators as dictated by logical functions. Here, we rely on the generalised framework initially introduced by R. Thomas and collaborators~\citep{thomas} and implemented in our software tool GINsim~\citep{chaouiya2012}.
Because precise knowledge of the durations of underlying mechanisms is often lacking, one assumes that, when multiple components are called to change their levels, all update orders have to be considered. This corresponds to the asynchronous updating scheme~\citep{thomas}. The dynamics of these models are classically represented by State Transition Graphs (STGs) where nodes embody the model states and edges represent the state transitions; each path in this finite, directed graph accounts for a potential trajectory of the system. 
These trajectories eventually end up in terminal strongly connected components that represent the model attractors.

Not surprisingly, the number of states of logical models grows exponentially with the number of regulatory components. Moreover, due to the asynchronous updating scheme, the dynamics are non-deterministic; they possibly encompass alternative trajectories towards a given state, as well as transient cycles. All this turns the analysis into a difficult challenge, particularly the identification of (reachable) attractors, which embody the long-term behaviours of the model.
In this context, methods have been developed to find point attractors -- or stable states -- and complex, oscillatory attractors (or, at least to, circumscribe their location)~\citep{naldi07b,garg08,zanudo13}. Here, we primarily aim at efficiently determining attractors reachable from specific initial condition(s) as well as estimating the reachability probability of each of those attractors. 

An STG is readily interpreted as the transition matrix of a (finite) Discrete-Time Markov Chain (DTMC).
Generally, STGs encompass distinct attractors (or recurrent classes) and thus define absorbing chains, while most existing results relate to recurrent (or irreducible) chains \citep{grinstead1997,levin2009markov,prum12a}.
Here, we focus on reachability properties and, more specifically, on absorption probabilities in finite DTMCs.
Moreover, we aim at avoiding the construction of the whole dynamics (or the associated transition matrix). To this end, we rely on the logical rules as an implicit description of the transition matrix.

Following a background section, we first introduce the \firefront algorithm, a quasi-exact method, which simultaneously follows all (concurrent) trajectories while propagating state probabilities. This algorithm follows a principle similar to those employed for infinite Markov chains \citep{munsky06,Henzinger09}.
Next, we present the \avatar algorithm, a Monte Carlo approach adapted to cope with the existence of strongly connected components in the trajectory.
Finally, in section \ref{sec:results}, both methods are applied to a range of models and compared to other tools, illustrating their respective performances and specificities.

\begin{methods}
\section{Methods}

In this section, we first briefly introduce the basics on logical regulatory graphs (LRGs), their state transition graphs (STGs) attractors as well as absorbing Markov chains. We then present \firefront, our first algorithm. The rest of the section focuses on \avatar, an adaptation of the classical Monte Carlo simulation to cope with cyclical behaviours. It is worth noting that, although for small enough LRGs, it is possible to explicitely construct  STGs and identify reachable attractors, it is not straightforward to evaluate reachability probabilities.

\subsection{Background}
\subsubsection{Basics on logical models and their dynamics}\label{sec:basics}

\begin{definition}\label{def:LRG}
A \emph{logical regulatory graph} (LRG) is a pair $(G,K)$, where:
\begin{itemize}
\item $G = \left\{ g_{i} \right\}_{i=0,\ldots n}$ is the set of regulatory components. Each $g_{i} \in G$ is associated to a variable $v_{i}$ denoting its level, which takes values in $D_{i} = \left\{0,\ldots M_{i}\right\} \subsetneq \mathbb{N}$;  $v = (v_{i})_{i=0,\ldots n}$ is a state of the system, and $S = \prod_{i=0,\ldots n} D_{i}$ denotes the state space. 
\item $(K_{i})_{i=0,\ldots n}$ denote the \emph{logical regulatory functions}; $K_{i} : S \rightarrow D_{i} $ is the function that specifies the evolution of $g_{i}$; $\forall v\in S,\,K_i(v)$ is the target value of $g_i$ that depends on the state $v$.
\end{itemize}
\end{definition}

The asynchronous dynamics of LRGs is represented by a graph as follows:

\begin{definition}\label{def:STG}
Given a logical regulatory graph $(G,K)$, its asynchronous state transition graph (STG) is denoted $(S,T)$, where:
\begin{itemize}
\item $S$ is the state space,
\item $T = \left\{(v,v') \in S^{2} \mid v' \in \Succ{v} \right\}$, 
where, for each state $v$, $\Succ{v}:S \rightarrow 2^S$ is the set of successor states $w$ such that: 
\end{itemize}
\begin{displaymath}
\exists g_{i} \in G \,\mbox{\rm with } \left\{
\begin{array}{l}
K_{i}(v) \not= v_{i} \,\mbox{\rm and }w_{i} = v_{i} + \frac{K_{i}(v)-v_{i}}{|K_{i}(v)-v_{i}|},   \\
\forall g_{j} \in G \setminus \{g_{i}\}, \quad w_{j} = v_{j}.\\
\end{array}
\right.
\end{displaymath}
\end{definition}

Note that, from the STG defined above, one can consider the  sub-graph reachable from a specific initial condition $v_0$.

We further introduce notation and classical notions that will be useful along the paper. 

We write $v \longrightarrow v'$ iff there exists a path between the states $v$ and $v'$ in a STG $(S,T)$. In other words, there is a sequence of states of $S$ such as: $s_0=v, s_1, \dots s_{k-1}, s_k=v'$, and for all $j\in\{1,\dots k\}$, $(s_{j-1}, s_j)\in T$.
We furthermore denote $v \overset{k}{\longrightarrow} v'$ such a path of length $k$. 
A strongly connected component (SCC) is a maximal set of states $A \subseteq S$ such that $\forall {v, v' \in A}$ with ${v \not= v'}$, $v\longrightarrow v'$. 
This is to say, there is a path between any two states in $A$ and this property cannot be preserved adding any other state to $A$.

Attractors of an LRG are defined as the {\it terminal} SCC of its STG (i.e., there is no transition leaving the SCC).
If it is a single state we call it a {\em point attractor}, otherwise it is a {\em complex attractor}.

\subsubsection{Markov chains and absorption}~\\\label{sec:MC}

The incidence matrix of a STG $(S,T)$ naturally translates into a transition matrix $\Pi$:
\begin{eqnarray*}
	\forall v,v'\in S && \Pi(v,v')>0 \Leftrightarrow (v,v')\in T,\\
	\forall v \in S && \Pi(v,v) = 1 \Leftrightarrow Succ(v) = \emptyset,\\
									&& \Pi(v,v) = 0 \,\, \mathrm{otherwise}.
\end{eqnarray*}

In this paper, probabilities of concurrent transitions are uniformly distributed: $\forall v\in S, \forall v'\in Succ(v),\, \Pi(v,v')=1/|Succ(v)|$. However, our methods  could be easily generalized.

 Given a STG  $(S,T)$, we consider its quotient graph with respect to the equivalence relation: $u \sim v \  \Leftrightarrow \  u \longrightarrow v \mbox{ and } v \longrightarrow u\,.$ 
 Let us now consider the Markov Chain on the finite state set $S$, defined by the initial law $\mu_0$ (that depends on the selection -- or not -- of an initial condition) and the transition matrix $\Pi$. In the quotient graph of $(S,T)$, each node gathers a set of states and corresponds to a class of the Markov chain.
The absorbing nodes of the quotient graph (i.e., nodes with no output arcs) form the absorbing classes of the chain, all the other classes being transient. 

Remark that the number of absorbing classes of the Markov chain  $(\mu_0, \Pi)$ is the number of attractors of the corresponding STG. Let $\theta$ be this number and $a_1, \dots, a_{\theta}$ the absorbing classes.

Now, let us stop the chain $(\mu_0, \Pi)$  when it reaches an absorbing class:  we thus define the Markov chain $X$ on the set $\tilde{S}= \TT\cup \AA$, where $\TT\subset S$ is the set of all the states belonging to a  transient class, and  
$\AA=\{ \{a_i\}, i =1, \dots \theta\}$ (each element $a_i$  being an absorbing class).
The transition matrix $\pi$ of $X$ is:
\begin{eqnarray*}
&& \pi(u,a_i)=\sum_{v \in a_i} \Pi(u,v) \quad  \forall u \in \TT, \forall a_i \in \AA\,,
\\
&& \pi(a_i,u)=0 \qquad \forall u \in \TT, \forall a_i \in \AA\,,
\\
&& \pi(a_i,a_i)=1 \qquad \forall a_i \in \AA\,,
\\
&& \pi(a_i,a_j)=0 \qquad \forall a_i \in \AA\,, \forall a_j \in \AA\,, i \neq j 
\\
&& \pi(u,v)= \Pi(u,v) \qquad \forall u,v \in \TT \,.
\end{eqnarray*}
If we reorder the states depending on whether they belong to $\TT$ or  to $\AA$, the transition matrix $\pi$ is a block matrix:
\begin{displaymath}
\pi= \left(
\begin{array}{cc}
Q & P
\\
0 & I
\end{array}
\right)\,,
\end{displaymath}
where $Q(u,v)=\pi(u,v)$ for $u,\,v \in \TT$, $P(u,a)=\pi(u,a)$ for $u\in \TT$ and $a \in \AA$, $0$ is the null matrix (no  transition from an absorbing class to a transient state), and $I$ the identity matrix.

Let us denote $\P_u(X_{k}=v) \stackrel{\Delta}{=}\P(X_{k}=v\,|\,X_0=u)=\pi^k(u,v)$ with 
 \begin{displaymath}
\pi^k = \left(
\begin{array}{cc}
Q^k & (\sum_{j=0}^{k-1} Q^j)\,P
\\
0 & I
\end{array}
\right)\,.
\end{displaymath}

Proofs of the next, well-known results can be found in e.g. \citep{grinstead1997}, chap. 11.

\begin{itemize}
\item $Q^k$ tends to 0 when $k$ tends to infinity, and
\begin{equation}\label{eq:limQ} \lim_{n \to +\infty} \sum_{k=0}^n Q^k = (Id-Q)^{-1}\,.\end{equation}
\item The hitting time of $\AA$ is almost-surely finite. 
\item From  any $u \in \TT$, the probability of $X$ being  absorbed in $a\in \AA$ is $\P_u(X_{\infty}=a)=(Id-Q)^{-1}\,P(u,a)\,.$
\end{itemize}

For simplicity, we will abuse terminology and refer to $\P_u(X_{\infty}=a)$ as the probability of  $a$.

\subsection{Firefront}

This is our first method to assess the reachability probability of the attractors.
Although simple, it is effective for restricted types of dynamics as demonstrated in section~\ref{sec:results}.
Briefly, the algorithm progresses in breadth from the initial state $v_0$, which is first assigned probability $1$.
The probabilities of each encountered state to all its successors are distributed and propagated, according to the transition matrix $\Pi$.  

At any given step $t$, the set of states being expanded and carrying a fraction of the original probability is called the \emph{firefront}, $F_t = \{v \in S, \exists v_0 \overset{t}{\longrightarrow} v\}$.
This procedure basically amounts to calculating, at each iteration $t$ and for each state $v$ the probability $\P_{v_0}(X_{t}=v)=\sum_{k=0}^{t} \pi^k(v_0,v)$.
In the context of $X$, the Markov chain introduced above, $\P_{v_0} (X_t \in F_t)=1$ and the firefront asymptotically   contains only states that are point attractors or members of complex attractors.


In practice, to avoid exploring unlikely paths, we introduce a new set of \emph{neglected states}, $N$.
Also, to ensure that the algorithm terminates whenever reachable attractors are all point attractors, we consider the set of \emph{attractors} $A$.
In the course of the exploration, the firefront, simply denoted $F$, will be reduced as explained below:
\begin{itemize}
	\item if the probability associated with a state $v \in F$ drops below a certain value $\alpha$, then $v$ is moved from $F$ to $N$.
As a consequence, the immediate successors of $v$ will not be explored at this time.
If a state $v \in N$ happens to accumulate more probability, as a result of being the successor of another state in $F$, and if this probability exceeds $\alpha$, then $v$ is moved from $N$ back to $F$;
	\item if a state in $F$ has no successors, it is moved to $A$;
if it is already in $A$, its probability increases according to this new incoming trajectory.
\end{itemize}

Importantly, the sum of probabilities of states in $F$, $N$ and $A$ is $1$.

Finally, the algorithm terminates when the total probability in $F$ drops below some predefined value $\beta$.
The probability thus associated to each point attractor $v \in A$ is a lower bound of $\P_{v_0}(X_{\infty}=v)$. An upper bound is obtained by adding to this value $\beta$ plus the total probability accumulated in $N$.
An outline of the \firefront algorithm is presented in Algorithm~\ref{alg:firefront}.

\begin{algorithm}\footnotesize
	\caption{\firefront algorithm \label{alg:firefront}}
\begin{algorithmic}[1]
\REQUIRE{$\alpha, \beta,  v_0$}
\ENSURE{$A$}

\STATE{$F \gets \{ v_0 \} \quad N \gets \emptyset \quad A \gets \emptyset$}

\WHILE{total probability in $F > \beta$}
	\STATE{$F' \gets \emptyset$}
	\WHILE{$F \not= \emptyset$}
		\STATE{$v \gets$ select and remove element of $F$}
		\IF{$\Succ{v} = \emptyset$}
			\STATE{$v$ is added to $A$ as a point attractor}
		\ELSE
			\FORALL{$v' \in \Succ{v}$}
				\STATE{$p \gets$ the probability of $v$ divided by $\left|\Succ{v}\right|$}
				\IF{$v'$ is in $F'$, $N$ or $A$}
				\STATE{Add $p$ to the probability of $v'$}
				\ELSE
				\STATE{Set the probability of $v'$ to $p$}
				\ENDIF
				\IF{probability of $v' \ge \alpha$}
				\STATE{Add $v'$ to $F'$ if it is not in $A$}
				\STATE{Remove $v'$ from $N$ if it is there}
				\ELSE
				\STATE{Add $v'$ to $N$}
				\ENDIF
			\ENDFOR
		\ENDIF
	\ENDWHILE
	\STATE $F \gets F'$
\ENDWHILE
\end{algorithmic}
\end{algorithm}

Unlike forest fires, which do not revisit burnt areas, the algorithm will, in general, revisit the same state in the presence of a cycle.
This would invalidate our colourful metaphor unless imagining uncannily rapid forest regeneration.
However, this poses some difficulty in the presence of complex attractors, because as described so far, the algorithm would never terminate.

Two modifications are introduced to obviate this problem. First, the algorithm is only allowed to perform a predefined maximum number of steps. Second, when available, the algorithm can be provided with a description of the complex attractors, effectively equipping \firefront with an \emph{oracle} deciding whether a state belongs to an enumerated complex attractor. In these cases, \firefront halts the exploration whenever it reaches a state recognized by the oracle, and treats all members of the corresponding attractor as a single element of $\mathcal{A}$ collectively accumulating incoming probability.

For Boolean models, the asymptotic running time of the algorithm is in $O\left(\log(n) \log^2(\alpha) n^{\lceil-\log(\alpha)\rceil}\right)$, where $n$ refers to the number of components of the model (see further details in the Supplementary Material). Additionally, the \firefront algorithm only needs to keep track of states in the firefront, $F$, the neglected set, $N$, and the identified attractors, $\mathcal{A}$, having otherwise no memory of the visited states, giving rise to a worst-case space complexity in $O(n^{\lceil-\log(\alpha)\rceil})$.


\subsection{Avatar}
We present the algorithm, termed \avatar, to identify the model attractors and assess their probability, considering the whole state space or specific initial condition(s). This adaptation of the classical Monte Carlo simulation aims at efficiently coping with (transient and terminal) SCCs.

\subsubsection{The algorithm}~\\
Monte Carlo simulation is classically used to estimate the likelihood of an outcome, when exhaustive enumeration is not feasible. For our problem, this means following random paths along the asynchronous dynamics (the STG). When, in the course of a simulation, a state with no successors is reached, this state is a point attractor. By performing a large number of simulations it is possible to evaluate the reachability probability of these point attractors. The simulation does not record past states, and thus memory requirements are minimal.
However, a major drawback is that it cannot identify cycles and thus complex attractors. Consequently, without restricting the maximum number of steps, the simulation does not terminate when a trajectory enters a terminal SCC. Moreover, in the presence of a transient cycle, it may re-visit the same states an unbounded number of times before exiting. For these reasons, we propose an appropriate modification of this approach.

\begin{algorithm}\footnotesize
\caption{The {\sc Avatar} algorithm (single simulation)\label{alg:avatar}}
\begin{algorithmic}[1]
\REQUIRE{$v_0$}
\ENSURE{$A$}
\STATE{$t \gets 0$ (first incarnation)}
\STATE{$v \gets v_0$}
\WHILE{$v$ has successors} 
	\STATE{$v' \gets$ random successor of $v$ taking into account transition probabilities}
	\IF{$v'$ was already visited in incarnation $t$}
		\STATE{$C^t \gets$ set of all states visited since the discovery of $v'$}
		\STATE{Rewire cycle $C^t$}
		\STATE{$t \gets t + 1$}
	\ENDIF
	\STATE{$v \gets v'$}
\ENDWHILE
\STATE{$C^{\ast}$ inductively defined by: \\
$C^{\ast}\gets \{v\}$ and $\forall w \in C^{\ast}, $ if $\exists k$ s.t. $ w \in C^{k}$ then $C^k \subseteq C^{\ast}$}
\STATE{$A \gets C^{\ast}$ (point if $|C^{\ast}| = 1$, complex otherwise)}
\end{algorithmic}
\end{algorithm}

\avatar is outlined in Algorithm~\ref{alg:avatar} (a detailed description of \avatar and its ancillary procedures is provided in the Supplementary Material). It  avoids  repeatedly visiting some states by detecting that a previously visited state is reached, indicating the presence of a cycle in the dynamics. Having detected a cycle, the algorithm proceeds by choosing one of the cycle exits. It is important, however, to choose this exit according to the probabilities of getting to each exit from the current state; the probability of a transition from any cycle state to a given exit must match the corresponding asymptotic probability, considering the infinitely many possible trajectories. The STG is thus rewired so as to replace all the transitions between the cycle states  by transitions from each cycle state  towards each cycle exit (see Fig.~\ref{ExModif}).  Each  rewiring creates a new so-called \emph{incarnation} of the dynamics. Such an incarnation --Sanskrit name of our algorithm-- is a graph with the same states as the original STG, but with different transition probabilities. This rewiring relies on theoretical fundamentals that are presented below. Upon rewiring, the simulation continues from the current state.

When a  state $v$ has no successors, we have to look to past incarnations to determine whether $v$ is a point attractor or is in a complex attractor. More precisely, if $v$ was ever part of a cycle in a previous incarnation, then $v$ belongs to an equivalence class containing all the states that have ever shared a cycle in past incarnations, with at least one of them having shared a cycle with $v$. 



As previously for \firefront, the algorithm can be complemented with an oracle that determines whether a state is part of an attractor. This obviously improves \avatar's performance. Moreover, \avatar not only evaluates the  probability of the attractors reachable from an initial condition, it  can also be used to assess the probability distribution of the attractors for the whole state space (i.e., considering all possible initial states).

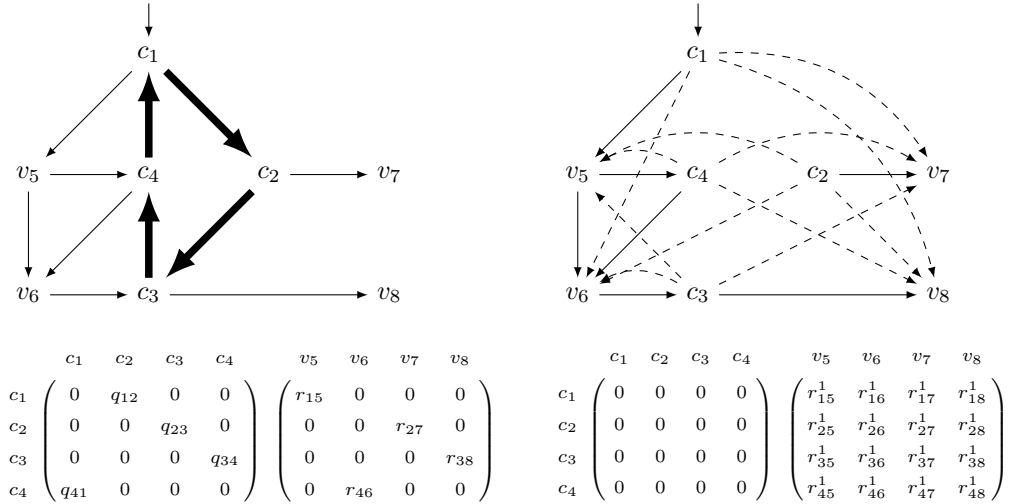
\begin{figure}
	\centering
\begin{tabular}{c@{\hspace{2.5cm}}c}
\begin{minipage}[c]{0.3\textwidth}
\begin{center}
\begin{tikzpicture}[scale=0.8]
\node (input) at  (2,7) {$$};
\node (c1) at  (2,6) {$c_1$};
\node (c2) at  (4,4) {$c_2$};
\node (c3) at  (2,2) {$c_3$};
\node (c4) at  (2,4) {$c_4$};
\node (c5) at  (0,4) {$v_5$};
\node (c6) at  (0,2) {$v_6$};
\node (c7) at  (6,4) {$v_7$};
\node (c8) at  (6,2) {$v_8$};
\draw[-latex ]  (input) to (c1);
\draw[-latex ]  (c1) to (c5);
\draw[-latex, line width=1mm ]  (c1) to (c2);
\draw[-latex, line width=1mm ]  (c2) to (c3);
\draw[-latex ]  (c2) to (c7);
\draw[-latex, line width=1mm ]  (c3) to (c4);
\draw[-latex ]  (c3) to (c8);
\draw[-latex, line width=1mm ]  (c4) to (c1);
\draw[-latex ]  (c4) to (c6);
\draw[-latex ]  (c5) to (c4);
\draw[-latex ]  (c5) to (c6);
\draw[-latex ]  (c6) to (c3);
\end{tikzpicture}
\end{center}
\end{minipage}
&
\begin{minipage}[c]{0.3\textwidth}
\begin{center}
\begin{tikzpicture}[scale=0.8]
\node (input) at  (2,7) {$$};
\node (c1) at  (2,6) {$c_1$};
\node (c2) at  (4,4) {$c_2$};
\node (c3) at  (2,2) {$c_3$};
\node (c4) at  (2,4) {$c_4$};
\node (c5) at  (0,4) {$v_5$};
\node (c6) at  (0,2) {$v_6$};
\node (c7) at  (6,4) {$v_7$};
\node (c8) at  (6,2) {$v_8$};
\draw[-latex ]  (input) to (c1);
\draw[-latex ]  (c1) to (c5);
\draw[-latex, dashed ]  (c1) to (c6);
\draw[-latex, dashed ]  (c1) to[bend left] (c7);
\draw[-latex, dashed ]  (c1) to[bend left] (c8);
\draw[-latex , dashed]  (c2) to[bend right] (c5);
\draw[-latex, dashed ]  (c2) to (c6);
\draw[-latex ]  (c2) to (c7);
\draw[-latex, dashed ]  (c2) to (c8);
\draw[-latex, dashed ]  (c3) to (c5);
\draw[-latex, dashed ]  (c3) to[bend right] (c6);
\draw[-latex, dashed ]  (c3) to (c7);
\draw[-latex ]  (c3) to (c8);
\draw[-latex, dashed ]  (c4) to[bend right] (c5);
\draw[-latex ]  (c4) to (c6);
\draw[-latex, dashed ]  (c4) to[bend left] (c7);
\draw[-latex, dashed ]  (c4) to (c8);
\draw[-latex ]  (c5) to (c4);
\draw[-latex ]  (c5) to (c6);
\draw[-latex ]  (c6) to (c3);
\end{tikzpicture}
\end{center}
\end{minipage}\\
\begin{minipage}[c]{0.3\textwidth}
\scriptsize{
$$
\bordermatrix{
              & c_1     & c_2     & c_3 & c_4     \cr
    c_1     & 0 & q_{12} & 0 & 0   \cr
    c_2     & 0 & 0 & q_{23} & 0     \cr
    c_3 & 0 & 0 & 0 & q_{34} \cr
    c_4     & q_{41} & 0 & 0 & 0     \cr
            }
\bordermatrix{
& v_5     & v_6     & v_7 & v_8     \cr
&r_{15} & 0 & 0 & 0\cr
&0 & 0 &r_{27} & 0\cr
& 0 & 0 & 0 & r_{38}\cr
&0 & r_{46}  & 0 & 0\cr
}$$
}\end{minipage}
&
\begin{minipage}[c]{0.3\textwidth}
\scriptsize{
$
\bordermatrix{
              & c_1     & c_2     & c_3 & c_4     \cr
    c_1     & 0 & 0 & 0 & 0   \cr
    c_2     & 0 & 0 & 0 & 0     \cr
    c_3 & 0 & 0 & 0 & 0 \cr
    c_4     & 0 & 0 & 0 & 0     \cr
            }
\bordermatrix{
& v_5     & v_6     & v_7 & v_8     \cr
&r^1_{15} & r^1_{16} & r^1_{17} & r^1_{18}\cr
&r^1_{25} & r^1_{26} &r^1_{27} & r^1_{28}\cr
&r^1_{35} & r^1_{36} &r^1_{37} & r^1_{38}\cr
&r^1_{45} & r^1_{46} &r^1_{47} & r^1_{48}\cr
}$
}
\end{minipage}
\end{tabular}

\caption{\label{ExModif}Modification of the transition matrix to force the random walk to exit a cycle. {\bf Left:} A subgraph, containing a cycle $(c_1,\,c_2,\,c_3,\,c_4)$ and its direct neighbours (or exit states) in the STG; below, the corresponding matrices $q$ and $r$. {\bf Right:} Modification as performed by \avatar: the cycle is dismantled and each of its states is connected to each exit state; below, the resulting matrices $q^1$ (null matrix) and $r^1$, with $r^1_{ij}=(Id-q)^{-1}r (v_i,v_j)$.}

\end{figure}

\subsubsection{Formal ground of \avatar rewiring}

The rewiring performed by \avatar in order to force the simulation to exit a cycle also modifies the probabilities associated to transitions. This is properly done so as to ensure a correct evaluation of the reachability probabilities following a (large) number of random walks over our Markov chain $X$. This procedure amounts to modifying the chain. Its rationale is formalised below and  illustrated in Fig.~\ref{ExModif}.

Suppose that $X_t=c_1$, and $X_{t+k}=c_1$ for $t$ and $k$ two positive integers. The walk has thus travelled along the cycle $C=(c_1,c_2,\dots c_k)$ (with $c_i\in S$ and  $(c_i,c_{i+1})\in T$, $\forall i=1,\dots k$). Note that this cycle may contain ''direct shorcuts``: $(c_i,c_j)\in T$, $j \neq i+1 \pmod{k}$.
We denote by $B$  the set of states directly reachable from $C$: $B=\{ v \in S\setminus C\,,\,(c_i,v)\in T, c_i \in C\}\,.$
Let $q$ be the $k \times k$ sub-matrix of $\pi$, for states $c_1,\dots c_k$, and $r$  the $k \times b$ sub-matrix of $\pi$ (with $|B|=b$), defining transitions from $C$ to $B$.
To force the walk to leave the cycle (rather than being trapped there for a long time), we locally modify  the transition matrix as follows:
\begin{itemize}
\item
 remove all the transitions (arcs) between the states of $C$; the sub-matrix $q$ is replaced by $q^1=0$, the null matrix;
\item
 add an arc from each state of $C$ to each state of $B$; 
the sub-matrix $r$ is replaced by $r^1 \stackrel{\Delta}{=}\sum_{j=0}^{\infty}q^j \,r$. By Eq.~(\ref{eq:limQ}), section \ref{sec:MC},

\mbox{$\forall c_i \in C,\,\forall v \in B,\,r^1(c_i,v)=\left[(Id-q)^{-1} r \right](c_i,v).
$}\end{itemize}

$Y$ denotes this new chain. Property \ref{prop:X=Y} asserts that, starting from any transient state $u$, $X$ and $Y$ have the same asymptotical behaviours.
\begin{property}\label{prop:X=Y}
 $\forall u \in \TT$,  $\forall a \in \AA$, $\P_u(Y_{\infty}=a) =  \P_u(X_{\infty}=a)\,.$
\end{property}
\begin{proof}Both chains have the same behaviour, except through the cycle: from $c_i$  the entry state in $C$, $X$  runs along the cycle during $l$ steps ($l \geq 0$) and may leave the cycle through a state $v\in B$ with transition probability $q^l r(c_i,v)$, whereas $Y$ would go directly from $c_i$ to $v$, with transition probability $r^1(c_i,v)$.  Therefore,  for all $ u \in \TT$, $ a \in \AA$ and $ j \geq 0, $ we have that $ 
\P_u(Y_{j}=a) \geq \P_u(X_{j}=a) $ and thus,
\begin{eqnarray*}
\sum_{j=1}^k  \P_u(Y_{j}=a) &\geq& \sum_{j=1}^k  \P_u(X_{j}=a)
\\
  \P_u(Y_{\infty}=a) &\geq&   \P_u(X_{\infty}=a)
\\
1=\sum_{a \in \AA}   \P_u(Y_{\infty}=a) &\geq& \sum_{a \in \AA} \P_u(X_{\infty}=a) = 1
\end{eqnarray*}
As all the terms being positive,  the Property is  proved. Therefore, \avatar and  Monte Carlo simulations  have asymptotically the same behaviours.
\end{proof}


\end{methods}

\section{Implementation}

Both \firefront and \avatar are implemented using Perl scripts, which rely on a common library of Perl modules. This enables a rapid prototyping of the algorithms and elicits a qualitative analysis of their performances. Both programs expect logical models to be specified in {\tt avatar} format, which is an export format produced by the current beta version of GINsim (available at \url{http://ginsim.org}). This format is human-readable and can be easily edited to specify initial conditions or known complex attractors (see program documentation for details).

Our implementation of \firefront halts the STG exploration after a predefined number of iterations (set by default to $|S|^2$). 

 \avatar implementation includes a heuristic optimisation controlled by parameters, whose default values generally work well. This optimisation considers  tradeoffs between costly rewirings and  simulations freely proceeding along cycles, and between memory cost of keeping state transitions after rewiring and not profiting from rewirings of previous simulations.

Since it is generally more efficient to rewire a larger transient than to perform multiple rewirings over portions of it, upon encountering a cycle,  \avatar performs an extension step controlled by a parameter $\tau$ that is a modified Tarjan's algorithm for SCC identification \citep{tarjan72} --  trajectories exploration is performed  up to a depth of $\tau$ away from states of the original cycle. 
The subsequent rewiring is then performed over the (potentially) extended cycle. In either case, \avatar does not rewire cycles having less than a predefined number of states. In the course of a single simulation, the value of $\tau$ is doubled upon each rewiring, in order to speed up the identification of  potential large transients. If, in further incarnations, the number of explicit state transitions exceeds a predefined number, the algorithm activates an \emph{inflationary mode} in which the  extension phase proceeds iteratively until an SCC is found.
This  mode is  always activated for models having an STG with less than a predefined number of states.
To alleviate the  cost of identifying and rewiring large transients, \avatar  keeps, for subsequent simulations, the explicit transitions corresponding to transient cycles involving more than a predefined number of states, provided these transients have an exit ratio (number of outgoing transitions per state) below~$1$.

\avatar  supports sampling over (portions of) the state space, in which case, each simulation   starts from a state randomly produced over the unconstrained components. 
When the sampling includes input components, attractors which are identical but for the values of sampled inputs are merged, and the different input valuations leading to each attractor are kept and reported.

Both \firefront and \avatar elicit the specification of oracles, which recognize subsets of the state space, specifically states of  known complex attractors. Conceptually, all the states recognized by an oracle are treated as if they were point attractors. In fact, whenever \avatar identifies a novel complex attractor in the course of a simulation, it   creates an oracle to be used in subsequent runs.

The output produced by both programs includes identified attractors and their estimated probabilities.
Optionally, graphics are produced with: (i) evolution of the number of states and cumulative probabilities in the firefront, neglected and attractor state sets for \firefront (see Fig.~\ref{fig:mcc:firefront}) and (ii) evolution of  estimated probabilities and trajectories length across all simulations for \avatar.

\begin{table}[h]
\centering
\resizebox{0.8\textwidth}{!}{
	\begin{tabular}{l|lllll}
	\toprule
	\multicolumn{1}{c|}{\textbf{Model name}} & \multicolumn{2}{c}{\textbf{\# Components}} & \multicolumn{2}{c}{\textbf{\# Attractors}} & \multicolumn{1}{c}{\textbf{State space size}}\\
		          & \textbf{Inputs} & \textbf{Proper}         & \textbf{Point} & \textbf{Complex} & \\
	\otoprule
	Random 1             & 0  & 10 & 1   & 1 & $1~024$\\
	Random 2             & 0  & 10 & 1   & 1 & $1~024$\\
	Random 3             & 0  & 15 & 1   & 1 & $32~768$\\
	Random 4             & 0  & 15 & 2   & 0 & $32~768$\\
	Synthetic  1          & 0  & 15 & 1   & 1 & $32~768$\\
	Synthetic  2          & 0  & 16 & 2   & 0 & $65~536$\\
	\otoprule
	Mammalian Cell Cycle       & 1  & 9  & 1   & 1 & $1~024$\\
	Segment Polarity (1-cell)  & 2  & 12 & 3   & 0 & $186~624$ \\
	Segment Polarity (2-cells) & 0  & 24 & 3   & 0 & $\approx9.7 \times 10^7$\\
	Segment Polarity (4-cells) & 0  & 48 & 15  & 0 & $\approx9.4 \times 10^{17}$\\
	Th differentiation  & 13 & 21 & 434 & 0 & $\approx3.9 \times 10^{10}$\\
	\bottomrule
	\end{tabular}
}
\caption{State-space characteristics of the models used as case studies.\label{tab:models}}
\end{table}

\section{Results}
\label{sec:results}

Here, we consider a set of randomly generated, synthetic and published biological models, briefly characterized below. 
We  analyse how \firefront and \avatar perform on these case studies and compare, when possible, to outcomes produced by BoolNet. This is an R package that not only generates random Boolean models, but also performs Markov chain simulations, identifying attractors \citep{boolnetR}.  We also briefly contrast \avatar and MaBoss, a related software, which  generalises Boolean models by defining  transition rates. MaBoss relies on Gillespie algorithm, it computes the temporal evolution of probability distributions and estimates stationary distributions \citep{maboss}.

\subsection{Case studies description}

Random models were generated using BoolNet, specifying the number of components and of regulators per component ~\citep{boolnetR}.
This process was automated in BoolNetR2GINsim, a small program available at \url{https://github.com/ptgm/BoolNetR2GINsim}, which accepts user-defined parameters, calls BoolNet and writes the resulting model to a GINML file (the GINsim  format). Generated models have 10 or 15 components,  each with  2 regulators and logical rules randomly selected (uniform distribution). From the resulting set of random models, we selected those exhibiting multi-stability (Table~\ref{tab:models}).

Additionally, we constructed a ''synthetic`` model exhibiting a very large complex attractor and few large transient cycles.
To further challenge our algorithms, we modified this model, adding one component in such a way that the complex attractor turned into a transient cycle with very few transitions leaving towards a point attractor (see synthetic models 1 and 2 in Table~\ref{tab:models}).

Our case studies also include published biological models.

The Boolean model of the mammalian cell cycle control ~\citep{faure2006} has 10  components and exhibits one point attractor (quiescent state) and one complex attractor (cell cycle progression). These attractors arise in (two) distinct disconnected regions of the state space, controlled by the value of the sole input component (CycD, which stands for the presence of growth factors).

 Sanchez {\it et al.}'s  multi-valued model of the segment polarity module --involved in  early segmentation of the {\it Drosophila} embryo-- defines an intra-cellular regulatory network,  then   instances of this network are connected  through inter-cellular signalling \citep{sanchez2008}. Here, we consider three cases:  the intra-cellular network (one cell),  the composition of two instances (i.e., two adjacent cells) and of four instances. Initial conditions are specified by the action of the pair-rule module (Wg-expressing cell for the single cell model) that operates earlier in development (see \cite{sanchez2008} for details).

The model for Th differentiation upon relevant environmental conditions embodied in  13 input components displays stable expression patterns that correspond to canonical Th1, Th2, Th17 and Treg subtypes and to hybrid cell types  \citep{naldi2010} .

\subsection{\firefront and \avatar in action}

All results are summarised in Table~\ref{tab:alg:results}.
\firefront  applied to random models 1, 2 and 3, identifies the point attractors and leaves the probability portion  related to the complex attractors undistributed, since it cannot identify complex attractors.
\avatar takes longer to identify and quantify all the attractors of random models 1, 2 and 3, but identifies the reachable complex attractors and reports each attractor's average depth.
For random model 4, \firefront needs 3.2 hours to perform 38 iterations, showing that  the running time is greatly influenced by the structure of the STG, which is explored  in a breath-first manner. Random model 4 has multiple large transient SCCs (data not shown), which causes \firefront to accumulate a large number of states in $F$. States of the transient SCCs are revisited until the probability of their incoming transitions drops below $\alpha$, which can take long.
\avatar on the other hand easily quantifies the two point attractors, despite the transient SCCs.
BoolNet is the fastest for these random models and MaBoss would give similar results.

For synthetic model 1, \firefront took 82 hours to  distribute the probability out of the large transient cycles, having performed the $10^3$ iterations allowed in our tests.
For synthetic model 2, \firefront could not distribute more than 6\% of the probability out of the transient SCC (purposely constructed with 8196 states and a dozen exits).
\avatar was able to escape the transient SCC due to its rewiring ability. It could identify and quantify the attractors for both models in less than 1 hour.
BoolNet on the other hand, completed synthetic models 1 and 2, after 7 and 5 days, respectively, which contrasts with its performance on the previous random models.
We also challenged MaBoss with these two models. The large complex attractor of synthetic model 1 could be found, provided an appropriate choice of the parameters, in particular the threshold for stationary distribution clustering \citep{maboss}. However, for synthetic model 2, no parametrisation could be found to prevent MaBoss identifying the transient SCC as a potential attractor.

\begin{figure}
	\centering
\includegraphics[width=.7\textwidth]{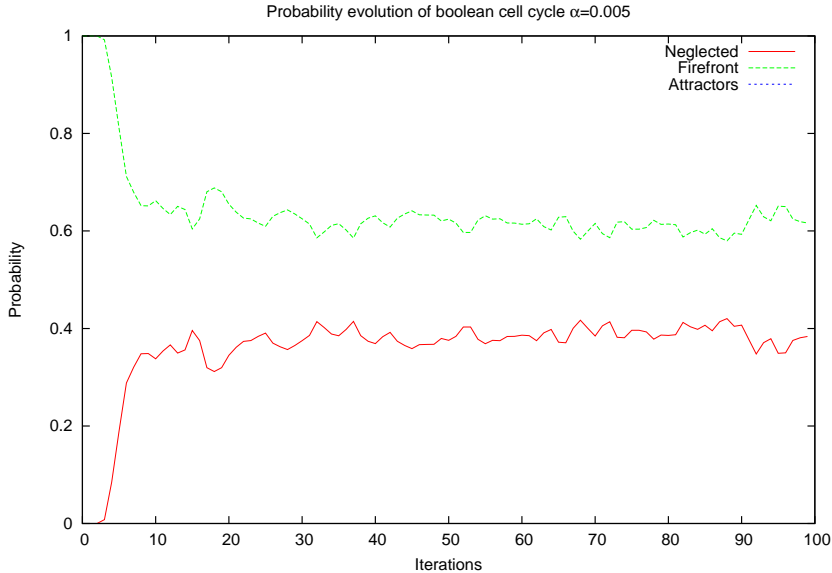}
\caption{Probability evolution of states  in the firefront, neglected and attractor sets produced by \firefront with the Mammalian Cell Cycle model, from an initial state leading the complex attractor and with $\alpha=0.005$ (for a better visualisation, we used a different value than the one used in Table \ref{tab:alg:results}). Note that the attractor set is empty; here only the complex attractor is reachable.\label{fig:mcc:firefront}}
\end{figure}

Starting in the region of the state space where the Mammalian Cell Cycle model  has a (unique) complex attractor, \avatar and BoolNet could assess it. Although \firefront could not identify the attractor, the  probabilities in the firefront and neglected sets sum up to 1, and the probability evolution stabilises (see  Fig.~\ref{fig:mcc:firefront}),  good indicators that a complex attractor has been found.
When sampling the state space,  \avatar and BoolNet could quantify it.

BoolNet was not applicable to the remaining multivalued models.

With the  Segment Polarity model, \firefront was rather fast for the single-cell case, despite the relatively large state space.
However, its performance quickly degrades for the two- and four-cell cases.
Since it did not reach the allowed maximum number of iterations, its stopping condition was that the total probability in $F$ felt below $\beta$, with all the residual probability in the neglected set, which in the end contains approximately 140, 52~000 and 210~000 states for the 1, 2 and 4 cell cases, respectively.
This indicates that $\alpha$ is not small enough with respect to the number of concurrent  trajectories towards the attractors.
Despite its worse running time for the single-cell case, \avatar's performance degrades much more slowly with successive model compositions.
Additionally, its single trajectory strategy allows it to move farther away from the initial state, thus identifying more point attractors.

Finally, we applied \avatar to the Th differentiation model, starting from an initial condition associated to the Th17 cell type, sampling a subset of input values (see Fig. 5 in \citep{naldi2010}).
By sampling over these initial states, \avatar identified the 4 reachable point attractors  (see Fig. 7 in \citep{naldi2010}).
Moreover, \avatar reported which input values led to each attractor (with the corresponding partial probability),  illustrating the usefulness of the algorithm for signalling-regulatory models.


\begin{table*}
\resizebox{\textwidth}{!}{
	\begin{tabular}{l l|clcc|ccc|cc}
	\toprule
	\multicolumn{1}{c}{\textbf{Name}} & \textbf{Initial} & \multicolumn{4}{c}{\textbf{\firefront} ($\alpha=10^{-5}$)} & \multicolumn{3}{c}{\textbf{\avatar} ($10^4$ runs)} & \multicolumn{2}{c}{\textbf{BoolNet} ($10^4$ runs)} \\
 & \textbf{conditions} & \textbf{Time} & \multicolumn{1}{c}{\textbf{Attractors} ($p$)}   & \multicolumn{1}{c}{\textbf{Residual}} & \textbf{Iterations} & \textbf{Time} &  \multicolumn{1}{c}{\textbf{Attractors} ($p$)} & \multicolumn{1}{c}{\textbf{Avg depth}} & \textbf{Time} & \multicolumn{1}{c}{\textbf{Attractors} ($p$)} \\
	\otoprule
	Random 1             & \emph{uncommitted}  & $57$s &
		{\begin{minipage}{1.1cm}
			\begin{tabular}{c@{~(}c@{)}}
			PA1 & 0.67
			\end{tabular}
		  \end{minipage} }
		& 0.33  & $10^3$ 
		& $12.4$min &
		{\begin{minipage}{1.1cm}
			\begin{tabular}{c@{~(}c@{)}}
			PA1 & 0.67\\
			CA2 & 0.33\\
			\end{tabular}
		  \end{minipage} }
		&   
				{\begin{minipage}{1.1cm}
			\begin{tabular}{c}
			9.18\\
			5.3\\
			\end{tabular}
		  \end{minipage} }
		&
		 $19$s &
			{\begin{minipage}{1.1cm}
			 \begin{tabular}{c@{~(}c@{)}}
			 PA1 & 0.67\\
			 CA2 & 0.33
			 \end{tabular}
			 \end{minipage} }	
	\\\midrule

	Random 2             & \emph{uncommitted}  & $2$s &
		 {\begin{minipage}{1.1cm}
			\begin{tabular}{c@{~(}c@{)}}
			PA1 & 0.25\\
			\end{tabular}
		  \end{minipage} }
			& 0.75  & $10^3$ 
			& $1.8$min &
			{\begin{minipage}{1.1cm}
			\begin{tabular}{c@{~(}c@{)}}
			PA1 & 0.25\\
			CA2 & 0.75
			\end{tabular}
		  \end{minipage} }
		  &  
		 	{\begin{minipage}{1.1cm}
			\begin{tabular}{c}
			6.43\\
			9.18\\
			\end{tabular}
		  \end{minipage} }
		&
		 $19$s &
			{\begin{minipage}{1.1cm}
			 \begin{tabular}{c@{~(}c@{)}}
			 PA1 & 0.25\\
			 CA2 & 0.75
			 \end{tabular}
			 \end{minipage} }	
	\\\midrule

	Random 3             & \emph{uncommitted}  & $30$s &
			{\begin{minipage}{1.1cm}
			\begin{tabular}{c@{~(}c@{)}}
			PA1 & 0.21\\
			\end{tabular}
		  \end{minipage} }	
			&  0.79 &  $10^3$ 
			& $5.3$min & 
			{\begin{minipage}{1.1cm}
			\begin{tabular}{c@{~(}c@{)}}
			PA1 & 0.21\\
			CA2 & 0.79\\
			\end{tabular}
		  \end{minipage} }
		  & 
		  {\begin{minipage}{1.1cm}
			\begin{tabular}{c}
			8.83\\
			8.45\\
			\end{tabular}
		  \end{minipage} } 
		&
		 $20$s &
			{\begin{minipage}{1.1cm}
			 \begin{tabular}{c@{~(}c@{)}}
			 PA1 & 0.20\\
			 CA2 & 0.80
			 \end{tabular}
			 \end{minipage} }	
	\\\midrule

	Random 4             & \emph{uncommitted}  & $3.2$h &
		 {\begin{minipage}{1.1cm}
			\begin{tabular}{c@{~(}c@{)}}
			PA1 & 0.40\\
			PA2 & 0.51\\
			\end{tabular}
		  \end{minipage} }	
		& 0.09 &  38  
		& $7.6$min &
		 {\begin{minipage}{1.1cm}
			\begin{tabular}{c@{~(}c@{)}}
			PA1 & 0.46\\
			PA2 & 0.54\\
			\end{tabular}
		  \end{minipage} }
		  &   
		  {\begin{minipage}{1.1cm}
			\begin{tabular}{c}
			20.64\\
			15.11\\
			\end{tabular}
		  \end{minipage} }
		&
		 $19$s &
			{\begin{minipage}{1.1cm}
			 \begin{tabular}{c@{~(}c@{)}}
			 PA1 & 0.46\\
			 PA2 & 0.54
			 \end{tabular}
			 \end{minipage} }	
	\\\midrule
	
	Synthetic  1   & \emph{uncommitted}  & $82$h &
		 {\begin{minipage}{1.1cm}
			\begin{tabular}{c@{~(}c@{)}}
			PA1 & 0.56\\
			\end{tabular}
		  \end{minipage} }	
		& 0.44 &  $10^3$  
	 & $35$min &
	  {\begin{minipage}{1.1cm}
		 \begin{tabular}{c@{~(}c@{)}}
		 PA1 & 0.58\\ 
		 CA1 & 0.42\\
		 \end{tabular}
		 \end{minipage} }
	   &
		 {\begin{minipage}{1.1cm}
			\begin{tabular}{c}
			18.45\\
			9.01\\
		  \end{tabular}
			\end{minipage} }
		&
		 $185.5$h &
			{\begin{minipage}{1.1cm}
			 \begin{tabular}{c@{~(}c@{)}}
			 PA1 & 0.60\\
			 CA2 & 0.40
			 \end{tabular}
			 \end{minipage} }	
	\\\midrule

	Synthetic  2  & \emph{uncommitted}  & $51.6$h &
		 {\begin{minipage}{1.1cm}
			\begin{tabular}{c@{~(}c@{)}}
			PA1 & 0.06\\
			PA2 & 10$^{-4}$\\
			\end{tabular}
		  \end{minipage} }	
		& 0.94 &  $10^3$  
	 & $58.5$min &
	  {\begin{minipage}{1.1cm}
		 \begin{tabular}{c@{~(}c@{)}}
		 PA1 & 0.07\\
		 PA2 & 0.93\\
		 \end{tabular}
		 \end{minipage} }
	   &
		 {\begin{minipage}{1.1cm}
			\begin{tabular}{c}
			27.15\\
			13.85\\
		  \end{tabular}
			\end{minipage} }
		&
		 $120$h &
			{\begin{minipage}{1.1cm}
			 \begin{tabular}{c@{~(}c@{)}}
			 PA1 & 0.08\\
			 PA2 & 0.92
			 \end{tabular}
			 \end{minipage} }	
	\\

	\otoprule
	 Mammalian Cell Cycle & $\mbox{CycD}=1$ & 
		 $2.08$min & 
		 {\begin{minipage}{1.1cm}
			\begin{tabular}{c@{~(}c@{)}}
			- - & 0.00\\
			\end{tabular}
		  \end{minipage} }
		 & $1.00$ & $10^3$ &
		$2.2$min &
		{\begin{minipage}{1.1cm}
			\begin{tabular}{c@{~(}c@{)}}
			CA1 & 1.00\\
			\end{tabular}
		  \end{minipage} }
	 		&
			{\begin{minipage}{1.1cm}
			\begin{tabular}{c}
			5.95\\
			\end{tabular}
		  \end{minipage} }
			&
			 $3.25$min & 
				{\begin{minipage}{1.1cm}
				\begin{tabular}{c@{~(}c@{)}}
				CA1 & 1.00\\
				\end{tabular}
				\end{minipage} }
		     \\\midrule
	 Mammalian Cell Cycle & \emph{sampling} 
	& \multicolumn{4}{c|}{N/A - due to sampling} &
	    $2.35$min &
	    {\begin{minipage}{1.1cm}
			\begin{tabular}{c@{~(}c@{)}}
			CA1 & 0.50\\
			PA2 & 0.50
			\end{tabular}
		  \end{minipage} }
	    & 
	    	{\begin{minipage}{1.1cm}
			\begin{tabular}{c}
			4.32\\
			2.76\\
			\end{tabular}
		  \end{minipage} }
			&
			 $1.83$min & 
				{\begin{minipage}{1.1cm}
				\begin{tabular}{c@{~(}c@{)}}
				CA1 & 0.50\\
				PA2 & 0.50
				\end{tabular}
				\end{minipage} }

			  \\\midrule
	Segment Polarity (1-cell)  & Wg-expressing cell  & $5$s &
			{\begin{minipage}{1.1cm}
			\begin{tabular}{c@{~(}c@{)}}
			PA1 & 0.84\\
			PA2 & 0.16\\
			\end{tabular}
		  \end{minipage} }
	 		& \textless$10^{-3}$ & 43 
			& $8.2$min &
			{\begin{minipage}{1.1cm}
			\begin{tabular}{c@{~(}c@{)}}
			PA1 & 0.84\\
			PA2 & 0.16\\
			\end{tabular}
		  \end{minipage} }
			  &
			{\begin{minipage}{1.1cm}
			\begin{tabular}{c}
			8.84\\
			11.17\\
			\end{tabular}
		  \end{minipage} }
			& \multicolumn{2}{c}{N/A - Boolean only}
			 \\\midrule

	Segment Polarity (2-cells) & Pair rule  & $17.2$h &
		  {\begin{minipage}{1.1cm}
			\begin{tabular}{c@{~(}c@{)}}
			PA1 & 0.65\\
			PA2 & 0.10\\
			\end{tabular}
		  \end{minipage} } & $0.25$ &  83 
		 	& $25.2$min &  
		  {\begin{minipage}{1.1cm}
			\begin{tabular}{c@{~(}c@{)}}
			PA1 & 0.89\\
			PA2 & 0.11\\
			PA3 & $10^{-4}$\\
			\end{tabular}
		  \end{minipage} }
		  & 
		  {\begin{minipage}{1.1cm}
			\begin{tabular}{c}
			38.83\\
			18.64\\
			49.00\\
			\end{tabular}
		  \end{minipage} }
			& \multicolumn{2}{c}{N/A - Boolean only}
			\\\midrule

	Segment Polarity (4-cells) & Pair rule  & $105.7$h &
		{\begin{minipage}{1.1cm}
			\begin{tabular}{c@{~(}c@{)}}
			PA1 & 0.13\\
			PA2 & 0.02\\
			PA3 & 0.01\\
			\end{tabular}
		  \end{minipage} } 
		 & 0.84 & 52 
		  & $1.2$h &
		  {\begin{minipage}{1.1cm}
			\begin{tabular}{c@{~(}c@{)}}
			PA1 & 0.87\\
			PA2 & 0.06\\
			PA3 & 0.06\\
			PA4 & 0.01\\
			PA5 & $10^{-3}$\\
			PA6 & $10^{-4}$\\
			PA7 & $10^{-4}$\\
			\end{tabular}
		  \end{minipage} }
		  &  
		  {\begin{minipage}{1.1cm}
			\begin{tabular}{c}
			59.12\\
			43.40\\
			36.51\\
			67.01\\
			55.10\\
			96.50\\
			138.00\\
			\end{tabular}
		  \end{minipage} }
			& \multicolumn{2}{c}{N/A - Boolean only}
		   \\\midrule

%
%

		Th differentiation reduced        & Th17+\emph{inputsampling} 
			 & \multicolumn{4}{c|}{N/A - due to sampling}
	 	& $1.5$min &
		  {\begin{minipage}{1.1cm}
			\begin{tabular}{c@{~(}c@{)}}
			PA1 & 0.63\\
			PA2 & 0.13\\
			PA3 & 0.12\\
			PA4 & 0.12\\
			\end{tabular}
		  \end{minipage} }
		  &   
		  {\begin{minipage}{1.1cm}
			\begin{tabular}{c}
			1.00\\
			7.00\\
			13.00\\
			4.00\\
			\end{tabular}
		  \end{minipage} }
			& \multicolumn{2}{c}{N/A - Boolean only}
			 \\

	\bottomrule	
\end{tabular}
}
\vspace{.3em}
\caption{Summary of the results for \firefront, \avatar  and BoolNet.
Parameters for  \firefront are $\alpha=10^{-5}$, maximum $10^3$  iterations; for \avatar the number of runs is $10^4$ (as well as BoolNet),   cycle extension parameters are the default ones.
For \firefront, residual probabilities are the sums of those found in firefront and neglected sets.
Point attractors are denoted PA and complex attractors CA.
Regarding  initial conditions, \textit{uncommitted} is a state in the basin of attraction of all the attractors and \textit{sampling} indicates a state randomly chosen at each run, where \textit{inputsampling} only samples over input components.
\label{tab:alg:results}}
\end{table*}

\section{Discussion and Conclusions}

 For models of biological networks, it is of a real interest to identify the  attractors reachable from initial conditions, as well as quantify their reachability.
Even though attractor identification could be achieved by keeping the STG in memory, this becomes infeasable for larger models due to combinatorial explosion.
In any case, we are still left with the problem of quantifying the reachability.

In this paper, we present two different strategies to address these problems.
\firefront, performs a memoryless breath-first exploration of the STG, avoiding any further exploration of states which fall below a given threshold $\alpha$. \avatar performs a modified version of the Monte Carlo algorithm, avoiding the exploration of states which have already been visited.

The best choice of algorithm and of optimal values for associated parameters would require  
information about the structure of the dynamics, which is generally unachievable.
The breadth of the explored STG as well as the structure  of transient SCCs clearly impacts on \firefront's performances. It is the degree of connectivity of  SCCs that influences  \avatar's performance.
Ideally, \avatar should avoid to rewire SCCs from which it can easily exit (low connectivity or high exit ratio). On the other hand, it should rewire SCCs from which it is hard to escape.
It is also much more efficient to rewire a whole SCC than to iteratively rewire portions of it.
While the size and structure of SCCs are not known {\it a priori},  \avatar incorporates heuristics that try to adapt the running parameters to the information collected in the course of the simulation.
Nevertheless, parameter tuning is warranted in cases not anticipated by the heuristics.

Unless provided with a so-called oracle, \firefront  does not detect complex attractors. Future work will address this limitation relying on the firefront content (stabilisation of its probability evolution as shown in Fig.~\ref{fig:mcc:firefront} and oscillation of its cardinal). 

Furthermore, we intend to integrate both algorithms into GINsim, our software tool devoted to logical modelling \citep{chaouiya2012}. We will also consider modified updating schemes and  non-uniform transition probabilities.

\section*{Acknowledgments}
\paragraph{Funding\textcolon}
This work was supported by Funda\c{c}\~ao para a Ci\^encia e a Tecnologia (FCT, Portugal), grants PTDC/EIA-CCO/099229/2008, PEst-OE/EEI/LA0021/2013 and IF/01333/2013.


\end{document}